\documentclass[11pt]{article}
\usepackage{amsmath}
\usepackage{graphicx}
\usepackage{amsfonts}
\usepackage{amssymb}
\usepackage{epsfig}
\usepackage{color}
\usepackage{psfrag}
\usepackage{epstopdf}
\usepackage{caption}
\usepackage{subcaption}

\newcommand{\EQ}{\begin{equation}}
\newcommand{\EN}{\end{equation}}
\newcommand{\be}{\begin{equation}}
\newcommand{\ee}{\end{equation}}
\newcommand{\bea}{\begin{eqnarray}}
\newcommand{\eea}{\end{eqnarray}}

\newcommand{\goto}{\rightarrow}

\def\goto{\longrightarrow}

\newcommand{\nn}{\nonumber \\}

\begin{document} \setcounter{page}{0}
\topmargin 0pt
\oddsidemargin 5mm
\renewcommand{\thefootnote}{\arabic{footnote}}
\newpage
\setcounter{page}{0}
\topmargin 0pt
\oddsidemargin 5mm
\renewcommand{\thefootnote}{\arabic{footnote}}
\newpage

\begin{titlepage}
\begin{flushright}
\end{flushright}
\vspace{0.5cm}
\begin{center}
{\large {\bf Critical points in the $CP^{N-1}$ model
}}\\
\vspace{1.8cm}
{\large Youness Diouane$^{1,2}$, Noel Lamsen$^{1}$ and Gesualdo Delfino$^{1}$}\\
\vspace{0.5cm}
{\em $^{1}$SISSA and INFN -- Via Bonomea 265, 34136 Trieste, Italy}\\
{\em $^{2}$ICTP, Strada Costiera 11, 34151 Trieste, Italy}\\

\end{center}
\vspace{1.2cm}

\renewcommand{\thefootnote}{\arabic{footnote}}
\setcounter{footnote}{0}

\begin{abstract}
\noindent
We use scale invariant scattering theory to obtain the exact equations determining the renormalization group fixed points of the two-dimensional $CP^{N-1}$ model, for $N$ real. Also due to special degeneracies at $N=2$ and 3, the space of solutions for $N\geq 2$ reduces to that of the $O(N^2-1)$ model, and accounts for a zero temperature critical point. For $N<2$ the space of solutions becomes larger than that of the $O(N^2-1)$ model, with the appearance of new branches of fixed points relevant for criticality in gases of intersecting loops.
 
\end{abstract}
\end{titlepage}

\newpage
\tableofcontents

\section{Introduction}
Determining if and how an additional local symmetry affects the universality class of a statistical model is a relevant issue in the theory of critical phenomena. A basic example is provided by the $RP^{N-1}$ model, in which $N$-component spin variables at each lattice site interact through an Hamiltonian invariant under global $O(N)$ rotations and local spin reversals. The local symmetry makes the difference with the usual $O(N)$ model and amounts to the head-tail symmetry characteristic of liquid crystals \cite{DgP}. In three dimensions, the weak first order transition observed in numerical simulations of the ferromagnetic model \cite{ZMZ} is consistent with the mean field scenario \cite{DgP}. On the other hand, in the two-dimensional case -- the one we focus on -- fluctuations are stronger and minimize the reliability of mean field predictions (see e.g. \cite{Cardy_book}), as illustrated by the phase transition of the three-state Potts model, which becomes continuous on planar lattices \cite{Wu}. For the $RP^{N-1}$ model, the absence of spontaneous breaking of continuous symmetry in two dimensions \cite{MWHC} generically suggests that criticality is limited to zero temperature, and numerical studies for $T\to 0$ show a fast growth of the correlation length which makes particularly hard to reach the asymptotic limit and draw conclusions about universality classes \cite{Sinclair,CEPS,CEPS2,NWS,Hasenbusch,CHHR,BFPV}. On the other hand, the possibility of finite temperature topological transitions similar to the Berezinskii-Kosterlitz-Thouless (BKT) one \cite{BKT} -- which should definitely occur for $RP^1\sim O(2)$ -- and mediated by "disclination" defects \cite{Stein,Mermin} has also been debated in numerical studies \cite{CPZ,KZ,FPB,DR,PFBo,FBBP,Tomita,SGR,KS}. While two-dimensional criticality has allowed for an impressive amount of exact solutions thanks to lattice integrability \cite{Baxter,Nienhuis} and conformal field theory \cite{BPZ,DfMS}, models with local symmetries traditionally remained outside the range of application of these methods. Recently, however, we showed in \cite{DDL,DLD} how the renormalization group fixed points of the $RP^{N-1}$ model can be accessed in an exact way in the scale invariant scattering framework \cite{paraf}, which implements in the basis of particle excitations the infinite-dimensional conformal symmetry characteristic of critical points in two dimensions and has provided in the last years new results for pure and disordered systems \cite{random,DT1,DT2,DL_ON1,DL_ON2,DL_vector_scalar, DL_softening} (see \cite{sis} for a review). We found that only $O(N(N+1)/2-1)$ fixed points exist for\footnote{See section~\ref{RPN} below for a question mark that we had left at $N=3$.} $N\geq 3$ and account for zero temperature criticality, while a line of fixed points yielding a BKT transition exists only for $N=2$. Our framework automatically yields results for $N$ real, corresponding to the known fact that lattice models such as $O(N)$ admit loop gas formulations \cite{Cardy_book,Nienhuis} in which $N$ plays the role of loop fugacity and does not need to be an integer. For $RP^{N-1}$ we found new branches of fixed points emerging below $N=2.24421..$ \cite{DLD}. 

In this paper we consider the basic lattice model with a continuous -- $U(1)$ -- local symmetry, namely the $CP^{N-1}$ model realized in terms of complex $N$-component spin vectors at lattice sites. In two dimensions, this model has been studied in the high energy context (since \cite{Eichenherr,DLDv,Witten}) for the similarities -- in particular asymptotic freedom -- which it shares with quantum chromodynamics, in statistical mechanics in relation with loop gases \cite{NSSO}, and in condensed matter in relation with quantum antiferromagnets (see e.g. \cite{TS}). The remarks that we made above for $RP^{N-1}$ concerning the continuous nature of the symmetry, zero temperature criticality, the possibility of topological transitions, and the absence of previous exact results, apply to $CP^{N-1}$ as well. We use scale invariant scattering to determine the exact fixed point equations for $CP^{N-1}$ symmetry, and find that the only solutions for $N\geq 2$ are of $O(N^2-1)$ type, also due to a special degeneracy emerging for $N=2,3$. This is consistent, in particular, with the known correspondence $CP^1\sim O(3)$. Our results again extend to real values of $N$, allowing us to see that quasi-long-range order and a BKT transition occur only for $N=\pm\sqrt{3}$, where $O(N^2-1)=O(2)$. Also here the space of solutions enlarges and new branches of fixed points appear below a threshold value of the symmetry parameter, which in this case turns out to be $N=2$. 

The paper is organized as follows. We recall the generalities of scale invariant scattering in section~2 and illustrate its application to the $O(M)$ model in section~3. Section~4 is devoted to the derivation of the fixed point equations for the $CP^{N-1}$ model, whose space of solutions is analyzed in section~5. The results for the $RP^{N-1}$ model are briefly recalled in section~6 for comparison, while the last section contains some final remarks. Two appendices complete the paper.

\section{Generalities of scale invariant scattering}
We begin our discussion by briefly recalling the generalities of scale invariant scattering \cite{paraf}, referring the reader to \cite{sis} for a review. The method relies on the fact that the continuum limit of a critical statistical system in two dimensions is described by a Euclidean field theory, which is the continuation to imaginary time of a quantum field theory defined in one space and one time dimension and exhibiting conformal invariance. In the quantum theory massless particles describe the excitations above the ground state (vacuum) and correspond to the fluctuation modes of the statistical system. Since in two dimensions conformal symmetry possesses infinitely many generators \cite{DfMS}, the scattering processes of the particles are subject to an infinite number of conservation laws, which force the final state to be kinematically identical to the initial one (completely elastic scattering). Moreover, scale invariance implies that the scattering amplitude of a two-particle process is a constant, namely does not depend on the center of mass energy, which is the only relativistic invariant and is dimensionful. 

\begin{figure}
     \begin{subfigure}[b]{.5\textwidth}
         \centering
         \includegraphics[width=.4\linewidth]{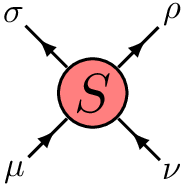}
     \end{subfigure}
     \hfill
     \begin{subfigure}[b]{.5\textwidth}
         \centering
         \includegraphics[width=.3\linewidth]{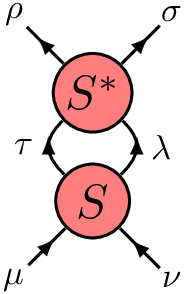}
     \end{subfigure}
\caption{Pictorial representations of the scattering amplitude $S_{\mu\nu}^{\rho\sigma}$ (left) and of the  product of amplitudes entering the unitarity equations~(\ref{unitarity}) (right).  
}
\label{scattering}
\end{figure}

These features are specific to two-dimensional criticality and substantially simplify the unitarity and crossing equations \cite{ELOP} that generally apply to relativistic scattering\footnote{See \cite{fpu} for a review on the off-critical regime in two dimensions.}. Let us denote by $\mu=1,2,\dots k$ the particle species, by $\mathbb{S}$ the scattering operator, and by $S_{\mu\nu}^{\rho\sigma}=\langle\rho\sigma|\mathbb{S}|\mu\nu\rangle$ the scattering amplitude for a process with particles $\mu$ and $\nu$ in the initial state and particles $\rho$ and $\sigma$ in the final state (figure~\ref{scattering}). Then the unitarity and crossing equations take the form \cite{paraf}
\begin{equation}
\sum_{\lambda,\tau} S_{\mu\nu}^{\lambda\tau}\left[S_{\lambda\tau}^{\rho\sigma}\right]^*=\delta_{\mu\rho}\delta_{\nu\sigma}\,,
\label{unitarity}
\end{equation}
\begin{equation}
S_{\mu\nu}^{\rho\sigma}=\left[S_{\mu\bar{\sigma}}^{\rho\bar{\nu}}\right]^*\,,
\label{cross}
\end{equation}
respectively\footnote{We denote by $\bar{\mu}$ the antiparticle of $\mu$.}. Invariance under charge conjugation, time reversal and spatial inversion provides the relations
\begin{equation}
S_{\mu\nu}^{\rho\sigma}=S_{\bar{\mu}\bar{\nu}}^{\bar{\rho}\bar{\sigma}}=S_{\nu\mu}^{\sigma\rho}=S_{\rho\sigma}^{\mu\nu}\,.
\label{cond}
\end{equation}

\section{Fixed points of the $O(M)$ model}
\label{OM}
\begin{figure}
\begin{center}
\includegraphics[width=8cm]{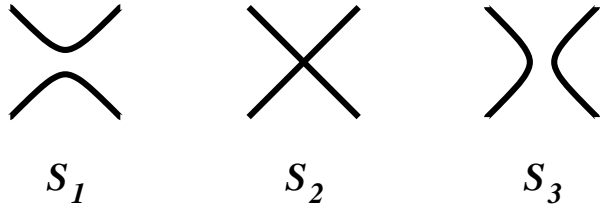}
\caption{Scattering amplitudes appearing in~(\ref{ON}); time runs upwards.
}
\label{vector_ampl}
\end{center} 
\end{figure}

Before turning to the $CP^{N-1}$ model it will be useful to briefly recall how scale invariant scattering applies to the $O(M)$ model \cite{paraf,DL_ON2}, which is defined on the lattice by the Hamiltonian
\EQ
{\cal H}_{O(M)} = -J\sum_{\langle i,j\rangle}{\bf s}_i\cdot{\bf s}_j\,,
\label{lattice_ON}
\EN
where ${\bf s}_i$ is a real $M$-component unit vector at site $i$ and the sum is taken over nearest neighbors. As usual, averages over configurations are performed with the Boltzmann weight $e^{-{\cal H}_{O(M)}/T}$, where $T$ is the temperature. The order parameter variable ${\bf s}_i$ corresponds in the scattering description to a vector multiplet of self-conjugated particles $a=1,2,\ldots,M$. The $O(M)$ tensorial structure involved in the scattering of a particle $a$ with a particle $b$ is preserved once the scattering matrix is written as
\begin{equation}
S_{ab}^{cd} = S_1\,\delta_{ab}\delta_{cd}+S_2\,\delta_{ac}\delta_{bd}+S_3\,\delta_{ad}\delta_{bc}\,,
\label{ON}
\end{equation}
with the amplitudes $S_1$, $S_2$ and $S_3$ accounting for annihilation, transmission and reflection, respectively (figure~\ref{vector_ampl}). The crossing equations (\ref{cross}) then yield
\bea
S_1=S_3^{*} &\equiv &  \rho_{1}\,e^{i\phi}, 
\label{cr1}\\
S_2 = S_2^* &\equiv & \rho_2,
\label{cr2}
\eea 
and lead to the parametrization of the amplitudes in terms of $\rho_2$ and $\phi$ real, and $\rho_1\geq 0$. It follows that the unitarity equations (\ref{unitarity}) can be written in the form
\bea
&& \rho_1^2+\rho_2^2=1\,, 
\label{u1}\\
&& \rho_1 \rho_2 \cos\phi=0\,,
\label{u2} \\
&& M \rho_1^2 + 2\rho_1\rho_2 \cos\phi +2\rho_1^2\cos2\phi=0\,. 
\label{u3} 
\eea

\begin{table}
\begin{center}
\begin{tabular}{l|c|c|c|c}
\hline 
Solution & $M$ & $\rho_1$ & $\rho_2$ & $\cos\phi$ \\ 
\hline \hline
I$_{\pm}$ & $(-\infty,\infty)$ & $0$ & $\pm 1$ & -  \\ 
II$_{\pm}$ & $[-2, 2]$ & $1$ & $0$ & $\pm\frac{1}{2}\sqrt{2-M}$  \\ 
III$_{\pm}$ & $2$ & $[0,1]$ & $\pm \sqrt{1- \rho_1^2}$ & $0$ \\[0.7em] 
\hline 
\end{tabular} 
\caption{Solutions of equations (\ref{u1})-(\ref{u3}), corresponding to the renormalization group fixed points with $O(M)$ symmetry.
}
\label{solutions}
\end{center}
\end{table}

\begin{figure}
\centering
\includegraphics[width=.65\textwidth]{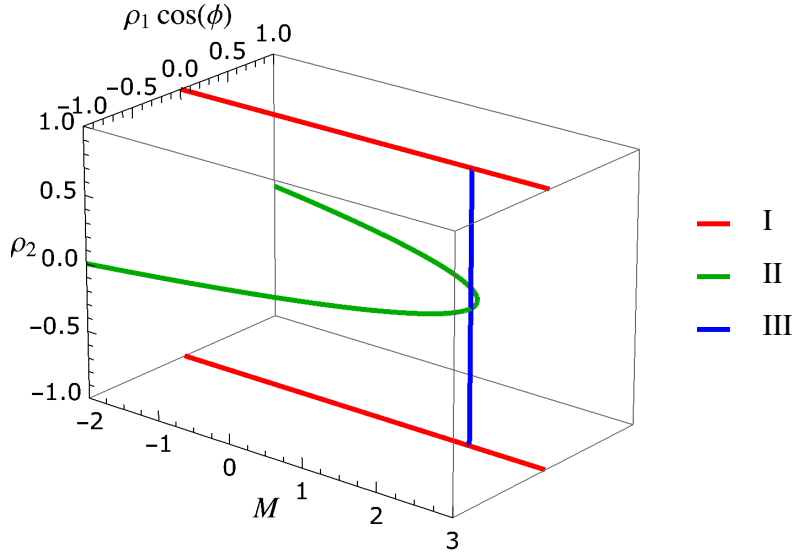}
\caption{Solutions of the $O(M)$ fixed point equations \eqref{u1}-\eqref{u3}. The two branches of II correspond to the critical lines for the dilute and dense phases of nonintersecting loops, III accounts for the BKT transition of the XY model, and the upper branch of I corresponds to the zero temperature critical point of the model for $M>2$.}
\label{OMfig}
\end{figure}

\noindent
Table~\ref{solutions} contains the solutions of equations (\ref{u1})-(\ref{u3}) \cite{paraf,DL_ON2} (also shown in figure~\ref{OMfig}), which yield the renormalization group fixed points with $O(M)$ symmetry. While a detailed discussion of the solutions is given in \cite{DL_ON2}, here we recall some basic features relevant for the remainder of the paper. The solutions II$_\pm$ are characterized by nonintersecting particle trajectories (namely $S_2=0$, see figure \ref{vector_ampl}), are defined in the range $M\in[-2,2]$, and meet at $M=2$. They are then identified as the critical lines of the dilute and dense regimes of the loop gas model, whose mapping on the partition function of the $O(M)$ model is well known \cite{Cardy_book,Nienhuis}. A particularly relevant feature of the loop formulation is that it implements on the lattice the continuation to noninteger values of $M$ that we see realized by equations (\ref{u1})-(\ref{u3}) directly in the continuum. The statistical properties of self-avoiding walks correspond to the limit $M\to 0$ \cite{DeGennes}. The correspondence between nonintersecting loop paths and nonintersecting particle trajectories was originally observed in \cite{Zamo_SAW} for the off-critical case.

The fact that solutions III$_\pm$, defined only for $M=2$, possess $\rho_1$ as a free parameter immediately identifies them with the line of fixed points at the origin of the BKT transition in the $XY$ model \cite{Cardy_book,BKT}. III$_+$ and III$_-$ meet at $\rho_1=1$, the BKT transition point where the field driving the transition is marginal \cite{paraf,DL_ON2}; it is instead irrelevant along III$_+$, so that this is the BKT phase in which correlations decay algebraically in the $XY$ model (quasi-long-range order). 

Finally, solutions I$_\pm$ are purely transmissive ($S_1=S_3=0$) and correspond to noninteracting bosons for $S_2=1$ and noninteracting fermions for $S_2=-1$. I$_+$ describes zero temperature criticality in the nonlinear sigma model with reduced Hamiltonian
\EQ
{\cal H}_{SM}= \frac{1}{T}\int d^2x\left(\nabla{\bf s}\right)^2, \hspace{1cm}{\bf s}^2=1\,,
\label{sigma}
\EN
where ${\bf s}(x)$ replaces in the continuum the lattice variable ${\bf s}_i$. The sigma model with $M>2$ describes the continuum limit of the $O(M)$ model in this range of $M$ and is characterized by exponentially diverging correlation length and vanishing interaction for $T\to 0$ (asymptotic freedom) \cite{Cardy_book,Zinn}. 

The solution I$_-$ yields a realization of the symmetry in terms of $M$ free fermions and is not relevant\footnote{Notice that, due to the quadratic nature of the unitarity equations (\ref{unitarity}), solutions differing for a change of sign of all amplitudes are always simultaneously present.\label{duplication}} for the critical behavior of the vector model (\ref{lattice_ON}) for generic $M$. The case $M=1$, however, allows some observations that will be useful in the subsequent sections. The symmetry $O(1)=\mathbb{Z}_2$ is that of the Ising model, which in two dimensions has a critical point described by a free neutral fermion \cite{DfMS}. The corresponding amplitude $S_{11}^{11}=-1$ is of course realized by I$_-$ in the purely transmissive form $S_{11}^{11}=S_2$. On the other hand, it is also realized by II$_-$ in the form $S_{11}^{11}=S_1+S_3$, as required by the fact that also the Ising partition function has a "geometrical" representation in terms of self-avoiding loops. This illustrates that a specific critical point may allow different diagrammatic realizations at the scattering level. Clearly, this is due to the fact that at $N=1$ there is a single particle species, and transmission, reflection and annihilation are not physically distinguishable\footnote{In a relativistic scattering process only the initial and final states are observable \cite{ELOP}.}. At the same time, some geometrical observables in the Ising model need to be computed in the limit $N\to 1$ \cite{DV_3point,DV_random}, and in this case solution II$_-$ provides the right analytic continuation.

\section{Fixed point equations of the $CP^{N-1}$ model}

The $CP^{N-1}$ lattice model is defined by the Hamiltonian
\begin{equation}
{\cal H}_{CP^{N-1}}=-J\sum_{\langle i,j \rangle}|\mathbf{s}_i \cdot {\mathbf{s}}_j^*|^2,
\label{cpham}
\end{equation}
where $\mathbf{s}_j$ is a $N$-component complex vector at site $j$ satisfying ${\bf s}_j\cdot{{\bf s}}_j^*=1$. The Hamiltonian \eqref{cpham} is invariant under global $U(N)$ transformations ($\mathbf{s}_j \to U \mathbf{s}_j$, $U\in U(N)$) and site-dependent $U(1)$ transformations ($\mathbf{s}_j \to e^{i \alpha_j} \mathbf{s}_j $, $\alpha_j\in\mathbb{R}$). These symmetries are represented through the tensorial order parameter variable
\begin{equation}
Q_i^{ab}=s_i^a({s}_i^b)^*-\frac{1}{N}\delta_{ab}\,.
\label{op}
\end{equation}
The presence of an invariant linear in the order parameter components is excluded by the constraint ${\bf s}_j\cdot{{\bf s}}_j^*=1$, which in turn makes $Q^{ab}_i$ traceless.

The implementation of scale invariant scattering for the two-dimensional $CP^{N-1}$ model at criticality proceeds through steps analogous to those seen in the previous section for the vector model. We first of all observe that in the continuum limit the order parameter field is now the Hermitian tensor $Q_{ab}(x)$, which creates particles that we label by $\mu=ab$, with $a$ and $b$ taking values from 1 to $N$. A state containing a particle $ab$ transforms under the $U(N)$ symmetry as 
\begin{equation}
|ab\rangle \goto |a'b'\rangle = \sum_{a,b} U_{a',a} U_{b',b}^* |ab\rangle\,,
\end{equation}
so that a scattering amplitude $S_{ab,cd}^{ef,gh}=\langle ef,gh|\mathbb{S}|ab,cd\rangle$ with particles $ab$ and $cd$ in the initial state and particles $ef$ and $gh$ in the final state transforms into
\begin{equation}
S_{a'b',c'd'}^{e'f',g'h'} = \sum_{a,b,c,d}\,\,\sum_{e,f,g,h} U_{a',a}U^*_{b',b} U_{c',c}U^*_{d',d}U^*_{e',e}U_{f',f}U^*_{g',g}U_{h',h} S_{ab,cd}^{ef,gh}\,.
\end{equation}
Taking also into account the relations \eqref{cond}, which can now be written as
\begin{equation}
S_{ab,cd}^{ef,gh} = S_{ba,dc}^{fe,hg}= S_{cd,ab}^{gh,ef} = S^{ab,cd}_{ef,gh}\,,
\end{equation}
$U(N)$-invariance corresponds to
\begin{figure}[t]
\centering
\includegraphics[scale=1.3]{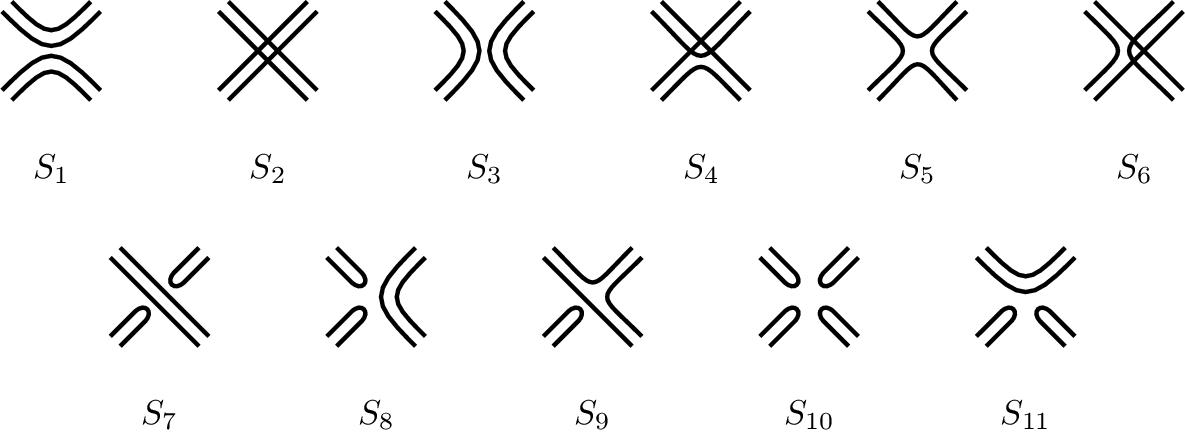}
\caption{Amplitudes entering~\eqref{Smatrix}. Time runs upwards}
\label{cpampl}
\end{figure}
\begin{equation}
\begin{split}
S_{ab,cd}^{ef,gh} &= S_1\, \delta _{a,d} \delta _{b,c} \delta _{e,h} \delta _{f,g} + S_2 \,\delta _{a,e} \delta _{b,f} \delta _{c,g} \delta _{d,h} + S_3 \,\delta_{a,g} \delta _{b,h} \delta _{c,e} \delta _{d,f} \\
   & \quad + S_4 \left(\delta
   _{a,d} \delta _{b,f} \delta _{c,g} \delta _{e,h}+\delta _{b,c}
   \delta _{a,e} \delta _{d,h} \delta _{f,g}\right) + S_5 \left(\delta
   _{b,c} \delta _{a,g} \delta _{d,f} \delta _{e,h}+\delta _{a,d}
   \delta _{b,h} \delta _{c,e} \delta _{f,g}\right)  \\
   & \quad + S_6 \left(\delta
   _{a,e} \delta _{b,h} \delta _{d,f} \delta _{c,g}+\delta _{b,f}
   \delta _{a,g} \delta _{c,e} \delta _{d,h}\right) + S_7 \left(\delta
   _{a,b} \delta _{e,f} \delta _{c,g} \delta _{d,h}+\delta _{c,d}
   \delta _{g,h} \delta _{a,e} \delta _{b,f}\right) \\
   & \quad + S_8 \left(\delta
   _{c,d} \delta _{e,f} \delta _{a,g} \delta _{b,h}+\delta _{a,b}
   \delta _{g,h} \delta _{c,e} \delta _{d,f}\right) + S_9 \big[\delta
   _{e,f} \left(\delta _{a,d} \delta _{b,h} \delta _{c,g}+\delta
   _{b,c} \delta _{a,g} \delta _{d,h}\right) \\
   & \qquad \qquad + \delta _{c,d}
   \left(\delta _{b,f} \delta _{a,g} \delta _{e,h}+\delta _{a,e}
   \delta _{b,h} \delta _{f,g}\right)  \delta _{a,b} \left(\delta
   _{d,f} \delta _{c,g} \delta _{e,h}+\delta _{c,e} \delta _{d,h}
   \delta _{f,g}\right) \\
   & \qquad \qquad + \delta _{g,h} \left(\delta _{a,d} \delta
   _{b,f} \delta _{c,e}+\delta _{b,c} \delta _{a,e} \delta
   _{d,f}\right)\big] + S_{10}\, \delta _{a,b} \delta _{c,d} \delta
   _{e,f} \delta _{g,h} \\
   & \quad+ S_{11} \left(\delta _{a,b} \delta _{c,d}
   \delta _{e,h} \delta _{f,g}+\delta _{e,f} \delta _{g,h} \delta
   _{a,d} \delta _{b,c}\right)\,,
\end{split} \label{Smatrix}
\end{equation}
with amplitudes $S_1,\dots,S_{11}$ depicted in figure~\ref{cpampl}. In this figure each incoming or outgoing particle has two terminals corresponding to its two indices, and a line connecting two indices corresponds to a Kronecker delta identifying them.

Crossing symmetry \eqref{cross} translates into
\begin{equation}
S_{ab,cd}^{ef,gh}=\left[S_{ab,hg}^{ef,dc}\right]^*\,.
\end{equation}
The crossing equations for the amplitudes $S_{i\leq 3}$ preserve the form (\ref{cr1}) and (\ref{cr2}), and we keep for these amplitudes the same parametrization in terms of $\rho_1$, $\rho_2$ and $\phi$. The crossing relations and the corresponding parametrizations for the remaining amplitudes are
\begin{align}
S_4&=S_6^*\equiv \rho_4 e^{i\theta},\label{cr_i}\\
S_5&=S_5^*\equiv\rho_5,\\
S_7&=S_7^*\equiv \rho_7,\\
S_8&=S_{11}^*\equiv \rho_8e^{i\psi},\\
S_9&=S_9^*\equiv\rho_9,\\
S_{10}&=S_{10}^*\equiv \rho_{10}\,,\label{cr_f}
\end{align}
with $\rho_5$, $\rho_7$, $\rho_9$, $\rho_{10}$, $\theta$ and $\psi$ real, and $\rho_4$ and $\rho_8$ nonnegative.

The unitarity condition \eqref{unitarity} can be written as 
\begin{equation}
\sum_{i,j=1}^N\sum_{k,l=1}^N S_{ab,cd}^{ij,kl}\left[S_{ij,kl}^{ef,gh}\right]^*=\delta_{a,e}\delta_{b,f}\delta_{c,g}\delta_{d,h}\,,
\label{unitaritytensor}
\end{equation}
and gives rise to the 11 independent equations
\begin{align}
1 &= \rho_1^2 + \rho_2^2 + 2 \rho_4^2, \label{eq:u1} \\
0 &= 2 \rho_1 \rho_2 \cos \phi + 2 \rho_4^2, \label{eq:u2}\\
0 &= N^2 \rho_1^2  + 2 \rho_1^2 \cos 2 \phi + 2 \rho_1 \rho_2 \cos \phi + 4 N \rho_1 \rho_4 \cos (\theta -\phi ) + 4 N \rho_1 \rho_5 \cos \phi \nn
&\qquad + 2 \rho_4^2 + 4 \rho_4 \rho_5 \cos \theta + 2 \rho_5^2 + 2 N \rho_1 \rho_8 \cos (\psi + \phi )  + 8 \rho_1 \rho_9 \cos \phi + 4 \rho_5 \rho_8 \cos \psi   \nn 
&\qquad + 8 \rho_4 \rho_8 \cos \theta  \cos \psi + 8 N \rho_8 \rho_9 \cos \psi +  N^2 \rho_8^2 + 8 \rho_9^2, \label{eq:u3}\\ 
0 &= 2 \rho_1 \rho_5 \cos \phi + 2 \rho_2 \rho_4 \cos \theta + N \rho_4^2 + N \rho_5^2 + 8 \rho_4 \rho_9 \cos \theta +  4 \rho_5 \rho_9  + 2 N \rho_9^2, \label{eq:u4} \\
0 &= 2 \rho_1 \rho_4 \cos (\theta +\phi ) + 2 \rho_2 \rho_5 + 2 N \rho_4 \rho_5 \cos \theta + 8 \rho_4 \rho_9 \cos \theta + 4 \rho_5 \rho_9 + 2 N \rho_9^2, \label{eq:u5} \\
0 &= 2 \rho_1 \rho_4 \cos (\theta -\phi ) + 2 \rho_2 \rho_4 \cos \theta, \label{eq:u6} \\
0 &= 2 \rho_1 \rho_8 \cos (\psi +\phi ) + 2 \rho_2 \rho_7 + 4 \rho_9 ( \rho_4 \cos \theta + \rho_7  + \rho_8 \cos \psi ) + N (\rho_7^2 +  \rho_8^2 + 2  \rho_9^2),  \label{eq:u7} \\
0 &= 2 \rho_1 \rho_7 \cos \phi + 2 \rho_2 \rho_8 \cos \psi + 4 \rho_9 ( \rho_4 \cos \theta + \rho_7 + \rho_8 \cos \psi) +   2 N ( \rho_7 \rho_8 \cos \psi + \rho_9^2), \label{eq:u8} \\
0 &= 2 \rho_1 \rho_9 \cos \phi + 2 \rho_2 \rho_9 + \rho_4^2 e^{-2 i \theta } + \rho_4 \rho_5  e^{-i \theta } + 2 \rho_4 \rho_7 \cos \theta + 2 \rho_4 \rho_8 e^{i \psi } \cos \theta \nn
&\qquad + N \rho_4 \rho_9 e^{-i \theta } + \rho_5 \rho_7 + \rho_5 \rho_8 e^{i \psi } + N \rho_5 \rho_9 + N \rho_7 \rho_9 + N \rho_8 \rho_9 e^{i \psi } + 4 \rho_9^2,  \label{eq:u9} \\
0 &= 2 \rho_1 \rho_{10} \cos \phi + 2\rho_2 \rho_{10} + 4 \rho_4 \rho_8 \cos (\theta -\psi ) + 8 \rho_7 \rho_8 \cos \psi + 4 N \rho_7 \rho_{10} + 6 N \rho_8 \rho_{10} \cos \psi \nn
&\qquad + 2 \rho_7^2 + 4 \rho_8^2 \cos 2 \psi + \left(N^2+2\right) \rho_8^2 + 8 N \rho_8 \rho_9 \cos \psi + 8 \rho_9^2 + 8 \rho_9 \rho_{10} + N^2 \rho_{10}^2, \label{eq:u10} \\
0 &= 2 \rho_1 \rho_4 e^{-i (\theta + \phi) } + 4 e^{-i \theta } \rho_4 \rho_9 + 2 e^{-i \theta } \rho_4 \rho_{10} + 2 e^{i \theta } \rho_4 \rho_{10} + N^2 \rho_8 \rho_{10} e^{i \psi } + N^2 \rho_1 \rho_8 e^{-i( \psi + \phi) } \nn
& \qquad + 2 N \rho_4 \rho_8 e^{-i (\theta + \psi) } + 2 N \rho_8^2 e^{2 i \psi } + 2 N \rho_5 \rho_8 e^{-i \psi  } + 2 N \rho_7 \rho_8 e^{i \psi } + 4 N \rho_1 \rho_9 e^{-i \phi } \nn
& \qquad + N \rho_1 \rho_{10} e^{-i \phi } + N \rho_8^2 + 4 N \rho_9 \rho_{10} + \rho_2 \rho_8 e^{-i\psi } + \rho_2 \rho_8 e^{i \psi } + 4 \rho_8 \rho_9 e^{-i \psi } + 8 \rho_8 \rho_9 e^{i \psi } \nn
& \qquad + \rho_1 \rho_8 e^{i (\phi -\psi) } + 3 \rho_1 \rho_8 e^{i( \psi - \phi) } + 2 \rho_1 \rho_7 e^{-i \phi } + 4 \rho_5 \rho_9+4 \rho_7 \rho_9 + 2 \rho_5 \rho_{10}\,. \label{eq:u11}
\end{align}
The choices of the indices yielding these equations are given in table \ref{choices}, where the notation $ab$ implies $a\neq b$; we checked that no new constraints arise from different choices.
\begin{table}
\centering
\begin{tabular}{|c|c|c|c|c|}
\hline
Equation & $\mu$ & $\nu$ & $\rho$ & $\sigma$\\
\hline\hline
\eqref{eq:u1} & $ab$ & $cd$ & $ab$ & $cd$\\
\hline
\eqref{eq:u2} & $ab$ & $cd$ & $cd$ & $ab$\\
\hline
\eqref{eq:u3} & $ab$ & $ba$ & $cd$ & $dc$\\
\hline
\eqref{eq:u4} & $ab$ & $bc$ & $ad$ & $dc$\\
\hline
\eqref{eq:u5} & $ab$ & $bc$ & $dc$ & $ad$\\
\hline
\eqref{eq:u6} & $ab$ & $cd$ & $ad$ & $cb$\\
\hline
\eqref{eq:u7} & $aa$ & $cd$ & $bb$ & $cd$\\
\hline
\eqref{eq:u8} & $aa$ & $cd$ & $cd$ & $bb$\\
\hline
\eqref{eq:u9} & $aa$ & $cd$ & $bd$ & $cb$\\
\hline
\eqref{eq:u10} & $aa$ & $bb$ & $dd$ & $cc$\\
\hline
\eqref{eq:u11} & $aa$ & $bb$ & $cd$ & $dc$\\
\hline         
\end{tabular}
\caption{External indices used in (\ref{unitarity}), \eqref{unitaritytensor} to obtain the unitarity equations \eqref{eq:u1}-\eqref{eq:u11}.}
\label{choices}
\end{table}

We still need to take into account that the field $Q_{ab}(x)$ that creates the particles is traceless. We do this requiring that  the trace mode  
\begin{equation}
{\cal{T}}=\sum_{a=1}^N aa
\end{equation}
does not interact with the generic particle $cd$ and can be discarded. This corresponds to 
\begin{equation}
\mathbb{S}|{\cal{T}} cd\rangle=S_0|{\cal{T}} cd\rangle \,, \hspace{1cm}S_0=\pm 1,
\label{eq:trdecoup}
\end{equation}
where the sign factor $S_0$ takes into account that the trace mode can decouple as a boson or a fermion. The last equation translates into $\sum_{a}S_{aa,cd}^{ef,gh} = S_0 \delta_{ef}\delta_{cg}\delta_{dh}$ and yields the relations
\begin{align}
S_0 &= \rho_2 + N \rho_7 + 2\rho_9, \\
0 &= \rho_1 e^{i \phi} + N \rho_8 e^{-i\psi} + 2\rho_9, \\
0 &= 2 \rho_4 \cos \theta + \rho_5 + N \rho_9, \\
0 &= \rho_7 + 2\rho_8 \cos \psi + N \rho_{10}\,.
\end{align}
These can be used to express $S_{i\geq7}$ in terms of $S_{i\leq 6}$ through
\begin{align}
\rho _7 &= \tfrac{1}{N} \left ( S_0 - \rho_2 + \tfrac{2}{N} \big ( 2 \rho_4 \cos \theta  + \rho_5 \big) \right ), \label{decoup1}\\
\rho _8 \cos \psi  &= \tfrac{1}{N} \left ( - \rho_1 \cos \phi + \tfrac{2}{N} \big ( 2 \rho_4 \cos \theta  + \rho _5 \big) \right ), \label{decoup2}\\
\rho _8 \sin \psi &= \tfrac{1}{N}\rho_1 \sin \phi, \label{decoup3}\\
\rho _9 &= -\tfrac{1}{N} \big ( 2 \rho_4 \cos \theta  + \rho_5 \big)  ,\label{decoup4}\\
\rho _{10} &= \tfrac{1}{N^2} \left ( 2 \rho_1 \cos \phi + \rho_2 - S_0  - \tfrac{6}{N} \big ( 2 \rho_4 \cos \theta  + \rho_5 \big) \right ). \label{decoup5}
\end{align}
When substituting \eqref{decoup1}-\eqref{decoup5} in \eqref{eq:u1}-\eqref{eq:u11}, the imaginary parts of \eqref{eq:u9} and \eqref{eq:u11} vanish, while their real parts as well as \eqref{eq:u7}, \eqref{eq:u8}, \eqref{eq:u10} become linear combinations of the first six equations. This reduces the unitarity equations \eqref{eq:u1}-\eqref{eq:u11} to six independent equations given by
\begin{align}
1 &= \rho_1^2 + \rho_2^2 + 2 \rho_4^2 \, , \label{uni1} \\
0 &= 2 \rho_1 \rho _2 \cos \phi + 2 \rho_4^2 \, , \label{uni2} \\
0 &= (N^2-1)\rho _1^2 + 2 \rho_1^2 \cos 2 \phi + 2 \rho_1 \rho_2 \cos \phi + 4 \left(N-\tfrac{1}{N}\right) \rho_1 \left ( \rho_4 \cos (\theta -\phi ) + \rho_5 \cos \phi  \right ) \nn
&\quad - \tfrac{4}{N} \rho_1 \rho_4 \cos (\theta +\phi )  + \tfrac{8}{N^2} \rho_4^2 \cos 2 \theta + 2 \left(1 + \tfrac{4}{N^2}\right) \rho_4 \left ( \rho_4 + 2 \rho_5 \cos \theta \right )\nonumber\\
&\quad + 2 \left(1+\tfrac{2}{N^2} \right) \rho_5^2 \, ,  \label{uni3} \\
0 &= 2 \rho_1 \rho_5 \cos \phi + 2 \rho_2 \rho_4 \cos \theta - \tfrac{4}{N} \rho_4^2 \cos 2 \theta + \left(N - \tfrac{4}{N}\right) \rho_4^2 - \tfrac{8}{N} \rho_4 \rho_5 \cos \theta \nonumber\\
&\quad + \left(N-\tfrac{2}{N}\right) \rho_5^2 \, , \label{uni4} \\
0 &= 2 \rho_1 \rho_4 \cos (\theta +\phi ) + 2 \rho_2 \rho_5  - \tfrac{4}{N} \rho_4^2 \cos 2 \theta - \tfrac{4}{N} \rho_4^2  + 2 \left(N-\tfrac{4}{N}\right) \rho_4 \rho_5 \cos \theta - \tfrac{2}{N} \rho_5^2 \, , \label{uni5} \\
0 &= 2 \rho _1 \rho _4 \cos (\theta -\phi )+2 \rho _2 \rho _4 \cos\theta\, . \label{uni6}
\end{align}
The solutions of these equations, which we discuss in the next section, correspond to the renormalization group fixed points with $CP^{N-1}$ symmetry in two dimensions. Notice that, since we derived the equations relying only on the symmetries of the Hamiltonian (\ref{cpham}), the space of solutions contains both the fixed points of the ferromagnetic case ($J>0$) and those of the  antiferromagnetic case ($J<0$). This point is explicitly illustrated in \cite{DT1,sis} for the case of the $q$-state Potts model.

\section{Solutions}
\begin{figure}[t]
\centering
\includegraphics[width=\textwidth]{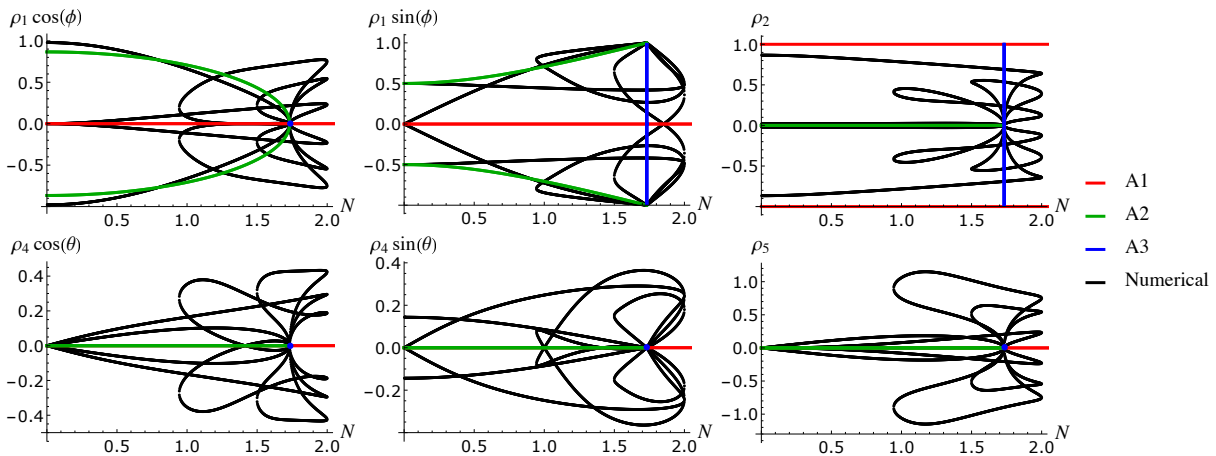}
\caption{Solutions of the $CP^{N-1}$ fixed point equations \eqref{uni1}-\eqref{uni6}.}
\label{cpfigmixsol}
\end{figure}

\begin{figure}
\centering
\includegraphics[width=0.7\textwidth]{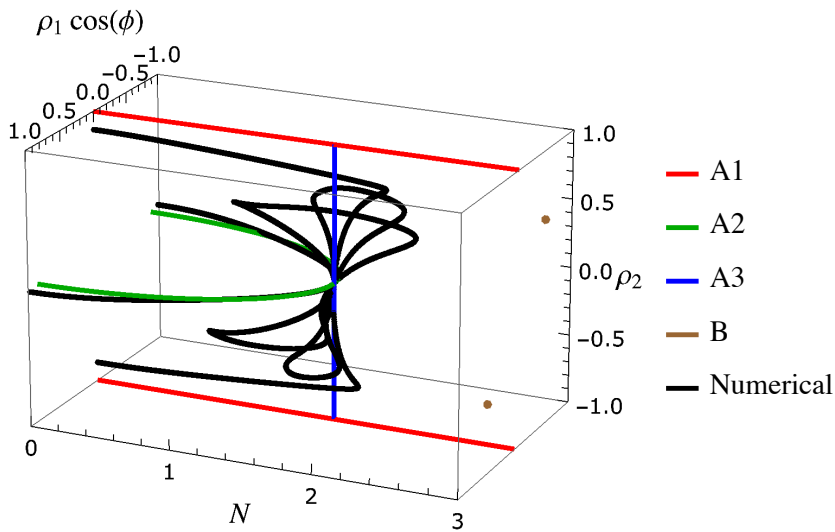}
\caption{Solutions of the $CP^{N-1}$ fixed point equations \eqref{uni1}-\eqref{uni6} in the parameter subspace $(\rho_1\cos\phi,\, \rho_2)$.}
\label{full}
\end{figure}

The solutions of the equations \eqref{uni1}-\eqref{uni6} that we determined analytically are listed in appendix~A and summarized in table \ref{sol}. The remaining solutions, which we dermined numerically for $N>0$, are shown in figure~\ref{cpfigmixsol} together with the analytical ones. The figure shows values of $N$ up to 2, since it turns out that only the solutions A1 and B exists beyond this value. Another visualization of the solutions is given in figure~\ref{full}. 

\begin{table}
\centering
\begin{tabular}{|c|c|c|c|c|c|c|c|}
\hline
Solutions & $N$ & $\rho_1$ & $\rho_2$ & $\cos\phi$ & $\rho_4$ & $\rho_5$ & $\cos\theta$\\
\hline\hline
A1$_\pm$ & $\mathbb{R}$ & $0$ & $\pm 1$ & $-$ & $0$ & $0$ & $-$\\
A2$_\pm$ & $[-\sqrt{3},\sqrt{3}]$ & $1$ & $0$ & $\pm\frac{1}{2}\sqrt{3-N^2}$ & $0$ & $0$ & $-$\\
A3$_\pm$ & $\pm \sqrt{3}$ &$\sqrt{1-\rho_2^2}$ & $[-1,1]$ & $0$ & $0$ & $0$ & $-$\\
B$_\pm$ & $3$ & $\frac{1}{2}$ & $\pm\frac{1}{2}$ & $\mp1$ & $\frac{1}{2}$ & $\rho_2$ & $\pm 1$\\
\hline
\end{tabular}
\caption{Analytic solutions of the $CP^{N-1}$ fixed point equations \eqref{uni1}-\eqref{uni6}.}
\label{sol}
\end{table}

We start the discussion of the solutions observing that when $\rho_4=\rho_5=0$ equations \eqref{uni1}-\eqref{uni6} reduce to the equations (\ref{u1})-(\ref{u3}) of the $O(M=N^2-1)$ model\footnote{$N^2-1$ is the number of independent real components of the order parameter variable (\ref{op}).}. As a consequence, the $CP^{N-1}$ model contains in particular the fixed points of the $O(N^2-1)$ model. This immediately allows to identify the solutions A1, A2 and A3 of table~\ref{sol} as corresponding to the $O(M=N^2-1)$ solutions I, II and III, respectively, of table~\ref{solutions}. The fact that $N^2-1=2$ when $N={\pm\sqrt{3}}$ explains the domain of definition of solutions A2 and A3. 

Since continuous symmetries do not break spontaneously in two dimensions \cite{MWHC}, the Hamiltonian (\ref{cpham}) is expected to possess only a zero temperature fixed point for $N\geq 2$. For $N>3$ we only have solution A1, which corresponds to an $O(N^2-1)$ fixed point\footnote{When discussing the Hamiltonian (\ref{cpham}) we refer to the bosonic realization A1$_+$ of the symmetry. The fermionic realization A1$_{-}$ is not relevant for that Hamiltonian.}. For $N=3$ the situation is apparently complicated by the existence of solution B. However, while solutions A1 and B clearly differ at the level of the amplitudes $S_1,\ldots,S_{11}$, it can be checked that they yield the same scattering matrix $(\ref{Smatrix})$. Hence, through the same mechanism we illustrated in section~\ref{OM} for the Ising model, the solutions A1 and B of the $CP^2$ model correspond to the same $O(8)$ fixed point\footnote{In three dimensions, where the symmetry can break spontaneously, a finite temperature critical point in the $O(8)$ universality class has been observed in numerical simulations of the antiferromagnetic $CP^2$ model \cite{DPV}.}. This is possible because for $N<4$ the particle indices do not take enough different values to make physically distinguishable all the terms entering the decomposition (\ref{Smatrix}).

Having clarified what happens for $N>2$, let us now consider $N=2$. Figures~\ref{cpfigmixsol} and \ref{full} show that $N=2$ is the value at which several pairs of solutions existing for $N<2$ meet and terminate. The list of solutions at $N=2$ is given in table~\ref{n2sol} in appendix~A. Such a proliferation is at first sight problematic, since we already argued that for $N\geq 2$ the Hamiltonian (\ref{cpham}) should possess only a zero temperature critical point. This is also fully consistent with the fact that $CP^1$ corresponds to the Riemann sphere, and then to $O(3)$. We can then suspect that, by the same mechanism observed for solution B at $N=3$, the solutions of table~\ref{n2sol} reconstruct the same scattering matrix (\ref{Smatrix}) than solution A1, and we checked that this is indeed the case. More specifically, solutions C3, C4, C7, C8, D3 and D4 correspond to A1$_+$, while C1, C2, C5, C6, D1 and D2 correspond to A1$_-$.

\begin{figure}[t]
\centering
\includegraphics[width=0.7\textwidth]{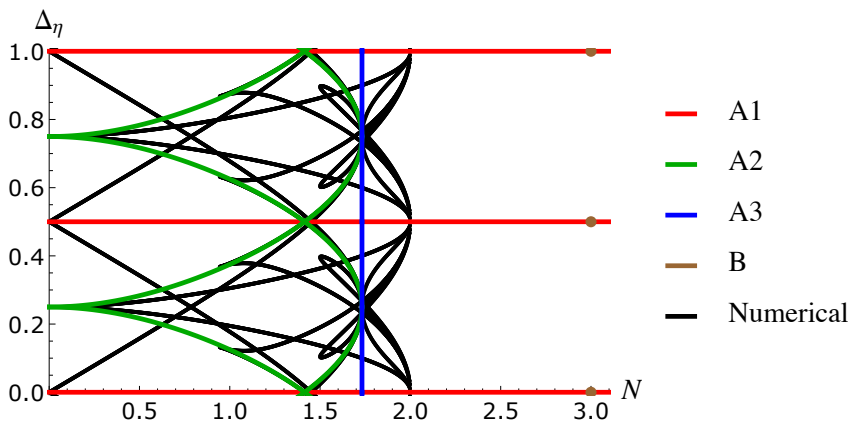}
\caption{The conformal dimension $\Delta_\eta$ for the different solutions of the $CP^{N-1}$ fixed point equations (\ref{uni1})-(\ref{uni6}).}
\label{confdim}
\end{figure}

We then see that for $N\geq 2$ all solutions of the $CP^{N-1}$ fixed point equations \eqref{uni1}-\eqref{uni6} correspond to $O(N^2-1)$ fixed points, and we already know that this is also the case for solutions A2 and A3. We finally need to consider the solutions that we determined numerically, which extend up to $N=2$, where they meet in pairs (see figures~\ref{cpfigmixsol} and \ref{full}). Since the meeting points at $N=2$ are $O(3)$ fixed points, and the $O(M)$ model does not possess branches of fixed points terminating at $M=3$, we can anticipate that the $CP^{N-1}$ branches terminating at $N=2$ correspond to new universality classes. We illustrate this fact considering the $U(N)$-invariant two-particle state
\begin{equation}
|\psi\rangle=\sum_{a,b=1}^{N}\big(|ab,ba\rangle-\frac{1}{N}|aa,bb\rangle\big)\,,
\label{eigenstate}
\end{equation}  
which scatters into itself, i.e. satisfies $\mathbb{S}|\psi\rangle=\lambda|\psi\rangle$, with an amplitude $\lambda$ which is a phase by unitarity and is given by
\begin{equation}
\lambda=(N^2-1)S_1+S_2+S_3+2\left(N-\frac{1}{N}\right)(S_4+S_5)-\frac{2}{N}S_6\,.
\label{lambda}
\end{equation}
Such a phase is related to the conformal dimension $\Delta_\eta$ of the chiral field that creates the particles as \cite{paraf,sis}
\begin{equation}
\lambda=e^{-2\pi i \Delta_\eta}.
\label{phase}
\end{equation} 
The values of $\Delta_\eta$ obtained through (\ref{lambda}) and (\ref{phase}) for the different solutions of the fixed point equations \eqref{uni1}-\eqref{uni6} are shown in figure \ref{confdim}. Equation (\ref{phase}) defines $\Delta_\eta$ modulo integers\footnote{This corresponds to the fact that in conformal field theory, given a primary field with dimension $\Delta$, there are descendants with dimension $\Delta+n$, $n=1,2,\ldots$\,. In addition, the duplication of solutions pointed out in footnote~\ref{duplication} causes $\Delta_\eta$ to go into itself under shifts by half-integers.}, and we plot the most relevant (in the renormalization group sense) interval $\Delta_\eta\in(0,1)$. The values 0 and 1/2 correspond to the $O(N^2-1)$ sigma model (solution A1$_{+}$) and to the fermionic realization (solution A1$_{-}$), respectively\footnote{See \cite{DL_ON2} for details about $\Delta_\eta$ in the $O(M)$ model.}. The figure clearly exhibits the collapse on the $O(N^2-1)$ solution A1 of the additional solutions existing at $N=2,3$. We also see that the numerical solutions at $N<2$ correspond to values of $\Delta_\eta$ -- and then to fixed points -- different from the $O(N^2-1)$ ones (A1, A2, A3). 

It appears from figure~\ref{cpfigmixsol} that the numerical solutions have nonvanishing $\rho_2$ and $\rho_4$. Hence, they correspond to intersecting particle trajectories (see figure~\ref{cpampl}) and should describe criticality in gases of intersecting loops. Actually, the relevance of $RP^{N-1}$ and $CP^{N-1}$ models for gases of intersecting loops was discussed in \cite{NSSO}. Here we are finding the corresponding $CP^{N-1}$ fixed points and showing that they exist up to $N=2$.

We then see that the fixed points of the $CP^{N-1}$ model coincide with those of the $O(N^2-1)$ model only for $N\geq 2$, where -- at least for $N$ integer -- the notion of continuous symmetry holds and does not allow for long range order. $O(N^2-1)$ fixed points are then obtained as a consequence of the fact that for $N\geq 2$ there are  only solutions with $\rho_4=\rho_5=0$ (or equivalent to them at $N=2,3$). When moving away from criticality, on the other hand, $\rho_4$ and $\rho_5$ are expected\footnote{Not for $N=2$, given that $CP^1\sim O(3)$.} to develop nonvanishing values, thus producing deviations\footnote{In particular, contrary to the $O(N^2-1)$ model \cite{ZZ}, the $CP^{N-1}$ model is not expected to be exactly solvable away from criticality \cite{BW,GW}.} from the off-critical $O(N^2-1)$ behavior that vanish as $T\to 0$. These conclusions parallel those we reached for the $RP^{N-1}$ model in \cite{DDL,DLD}, whose basic findings we recall in the next section.

\section{Parallels with the $RP^{N-1}$ model}
\label{RPN}
\begin{figure}[t]
\centering
\includegraphics[width=\textwidth]{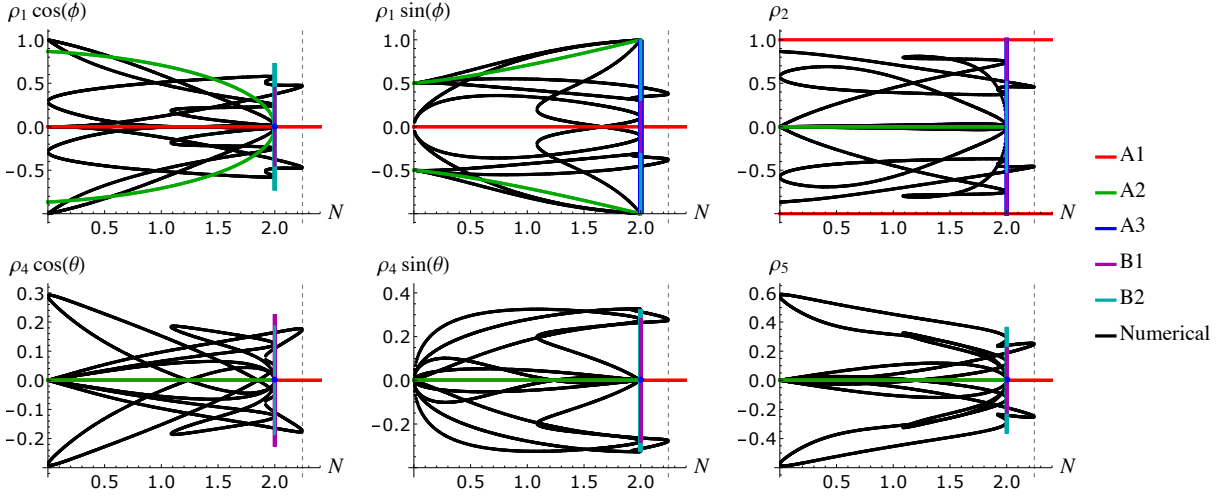}
\caption{Solutions of the fixed points equations for $RP^{N-1}$ symmetry \cite{DDL,DLD}. The dashed vertical line indicates the value $N=2.24421..$\,.}
\label{fignum}
\end{figure}

We briefly point out similarities and differences between the above results for the $CP^{N-1}$ model and those obtained for the $RP^{N-1}$ model in refs. \cite{DDL,DLD}, to which we refer the reader for the detailed derivation. The $RP^{N-1}$ model, defined by the lattice Hamiltonian
\begin{equation}
{\cal H}_{RP^{N-1}}=-J\sum_{\langle i,j \rangle}(\mathbf{s}_i \cdot {\mathbf{s}}_j)^2,
\label{rpham}
\end{equation}
differs from $CP^{N-1}$ for the fact that the spin variable ${\bf s}_i$ is real, so that the model is invariant under global $O(N)$ transformations and local spin inversions. The scale invariant scattering description proceeds through steps analogous to those of the present paper, starting from an order parameter that is now a traceless symmetric tensor. This allows a larger number of contractions between pairs of particle indices, but there are still 11 amplitudes $S_1,\ldots,S_{11}$ parametrized as in (\ref{cr1}), (\ref{cr2}), (\ref{cr_i})-(\ref{cr_f}). When $\rho_4=\rho_5=0$, the fixed point equations reduce to those of the $O(M_N)$ model, with $M_N=N(N+1)/2-1$. As a consequence, there are solutions A1, A2 and A3 that correspond to the solutions I, II and III, respectively, of table~\ref{solutions} with $M=M_N$. A1 is the only solution for $N>2.24421..$\,. More precisely, at $N=3$ there is an isolated solution B3, but we have now checked that it is equivalent to A1 by the same mechanism discussed in the previous section for solution B in $CP^2$. At $N=2$, solution A3 goes along with two additional solutions, B1 and B2, which also possess a free parameter and provide alternative realizations of the BKT phase in the $RP^1\sim O(2)$ model. Finally, we show in figure~\ref{fignum} how for $N<2.24421..$ there is a rich pattern of solutions that we determined numerically. 

\begin{figure}[t]
\centering
\includegraphics[width=0.8\textwidth]{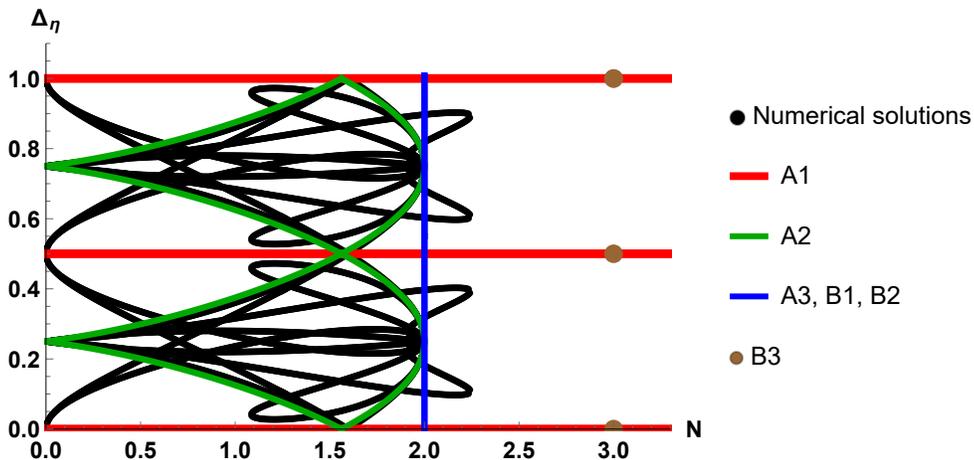}
\caption{$\Delta_\eta$ for the different $RP^{N-1}$ solutions.}
\label{Delta_RP}
\end{figure}

We can again use (\ref{phase}) to determine the conformal dimension $\Delta_\eta$, taking into account that (\ref{lambda}) is now replaced by 
\EQ
 \tfrac{(N-1)(N+2)}{2}S_1 + S_2 + S_3 + 2\tfrac{(N-1)(N+2)}{N} (S_4 + S_5) + 2\tfrac{N-2}{N}S_6\,,
\EN
in terms of the $RP^{N-1}$ amplitudes $S_i$ given in \cite{DDL,DLD}. The result for the different solutions is shown in figure~\ref{Delta_RP}. Notice that $N=2.24421..$ is the threshold value below which solutions that are not (or not equivalent to) $O(M)$ solutions appear, a threshold that in $CP^{N-1}$ occurs at $N=2$. While in the previous section the correspondence $CP^1\sim O(3)$ allowed us to anticipate that all the "threshold solutions" should be equivalent to A1, a similar argument is absent at the $RP^{N-1}$ threshold, and indeed figure~\ref{Delta_RP} illustrates that the solutions at $N=2.24421..$ are not related to A1.

\section{Conclusion}
In this paper we used scale invariant scattering theory to determine the exact fixed point equations of the two-dimensional $CP^{N-1}$ model for real values of $N$. We found that only solutions of $O(N^2-1)$ type exist for $N\geq 2$, and account for a zero temperature critical point. Additional solutions existing at $N=2,3$ actually correspond to alternative scattering realizations of the same $O(N^2-1)$ fixed points. We also found that a topological transition of BKT type only exists for $N=\pm\sqrt{3}$, where $O(N^2-1)=O(2)$. Several branches of fixed points that are not of $O(N^2-1)$ type appear below $N=2$ and are expected to describe criticality in loop gases. These solutions are characterized by amplitudes with nonvanishing transmission and correspond to gases of intersecting loops, examples of which were considered in \cite{NSSO} and related to the $RP^{N-1}$ and $CP^{N-1}$ models. It is interesting to observe how these new fixed points only emerge for $N<2$, where the model only makes sense within the loop gas continuation to real values of $N$. For integer $N\geq 2$, the continuous symmetry does not allow ordered phases in two dimensions, and the fixed point equations have solutions only in the subspace with $\rho_4=\rho_5=0$, where they coincide with those of the $O(N^2-1)$ model. On the other hand, for $N>2$, these scattering parameters are expected to develop nonzero values, so that the $O(N^2-1)$ behavior only arises asymptotically in the zero temperature limit. An exception is provided by $N=2$, since $CP^1\sim O(3)$. We recalled for comparison how in the $RP^{N-1}$ model, where $RP^1\sim O(2)$ has a finite temperature BKT transition, the space of fixed point solutions becomes larger than that of the $O(N(N+1)/2-1)$ model for $N\leq 2.24421..$ \cite{DDL,DLD}. 

It is also worth stressing that scale invariant scattering theory only exploits conformal invariance of critical points and the internal symmetry of the Hamiltonian, so that the space of solutions of the fixed point equations includes the critical points for both the ferromagnetic and the antiferromagnetic cases, a circumstance illustrated in more detail in \cite{DT1,sis} for the $q$-state Potts model.

We finally point out that our results for the fixed points of the renormalization group, characterized by diverging correlation length and scale invariance, add nothing to the debate \cite{SS,TS,CMP} about the possibility of a first order transition for $N$ large in $CP^{N-1}$ and $RP^{N-1}$ models, for which a first order transition at $N=\infty$ was deduced in \cite{MR,KZ2} and shown to be absent in numerical simulations performed up to $N=40$ \cite{KZ}.

\appendix
\section{Analytic solutions}

We list in this appendix the solutions of the fixed point equations (\ref{uni1})-(\ref{uni6}) that we determined analytically. With respect to table \ref{sol}, we also use the equations \eqref{decoup1}-\eqref{decoup5} to express the amplitudes $S_{i\geq 7}$.

\noindent $\bullet$ Solution A1a$_{\pm}$ is defined for $N\in \mathbb{R}$  and reads
\begin{equation}
\begin{split}
&\rho_2=S_0 \, \, , \, \, \rho_1=\rho_4=\rho_5=\rho_8=\rho_7=\rho_9=\rho_{10}=0\,. \label{sol11}
\end{split}
\end{equation}
\noindent $\bullet$ Solution A1b$_{\pm}$ is defined for $N\in\mathbb{R}$ and reads
\begin{equation}
\begin{split}
&\rho_2=-S_0 \, \, , \, \, \rho_1=\rho_4=\rho_5=\rho_8=\rho_9=0\,,\\
&\rho_7=\frac{2S_0}{N} \, \, , \, \, \rho_{10}=-\frac{\rho_7}{N}, \label{sol12}
\end{split}
\end{equation}
\noindent $\bullet$ Solution A2$_\pm$ is defined for $N\in [-\sqrt{3},\sqrt{3}]$ and reads\footnote{Signs enclosed in parenthesis are both allowed.}
\begin{equation}
\begin{split}
&\rho_1=1 \, \, , \, \, \rho_2=\rho_4=\rho_5=\rho_9=0 \, \, , \, \, \cos\phi=(\pm)\frac{1}{2}\sqrt{3-N^2} \,,\\
&\sin\phi=(\pm)\frac{1}{2}\sqrt{1+N^2} \, \, , \, \, \rho_7=\frac{S_0}{N} \, \, , \, \, \rho_8=\frac{1}{|N|} \, \, , \, \, \cos\psi=-\text{sgn}(N)\cos\phi,\\
&\sin\psi=\text{sgn}(N)\sin\phi \, \, , \, \, \rho_{10}=\frac{2\cos\phi}{N^2}-\frac{S_0}{N^2}\,.\label{sol2}
\end{split}
\end{equation}
\noindent$\bullet$ Solution A3$_\pm$ is defined for $N=\pm \sqrt{3}$ and reads
\begin{equation}
\begin{split}
&\rho_1=\sqrt{1-\rho_2^2} \, \, , \, \, \rho_2\in[-1,1] \, \, , \, \, \rho_4=\rho_5=\rho_9=0 \, \, , \, \, \phi=(\pm)\frac{\pi}{2} \, \, , \, \, \psi=\pm\phi,\\
&\rho_8=\frac{1}{|N|}\sqrt{1-\rho_2^2} \, \, , \, \, \rho_7=\frac{S_0-\rho_2}{N} \, \, , \, \, \rho_{10}=-\frac{\rho_7}{N}\,.\label{sol3}
\end{split}
\end{equation}
\noindent $\bullet$ Solution B$_\pm$ is defined for $N=3$ and reads
\begin{equation}
\begin{split}
&\rho_1=\rho_4=\rho_8=\frac{1}{2} \, \, , \, \, \rho_2=\rho_5=\rho_9=\pm \frac{1}{2} \, \, , \, \,  \phi=\frac{\pi}{2}\pm \frac{\pi}{2}=\theta+\pi=\psi+\pi,\\
&\rho_7=\frac{\rho_2+S_0}{3} \, \, , \, \, \rho_{10}=-\frac{\rho_7}{3}\mp \frac{1}{3}\,.\label{sol4}
\end{split}
\end{equation}

In the next appendix we show that solutions (\ref{sol11}) and (\ref{sol12}) differ only for the way the trace mode decouples (as a free fermion or a free boson); this is why they both appear in table~\ref{sol} as solution A1. Table~\ref{n2sol} gives the solutions at $N=2$.

\begin{table}
\hspace{-2.5cm}
{\small
\begin{tabular}{|c|c|c|c|c|c|c|}
\hline
Solution &$\rho_1\cos\phi$ & $\rho_1\sin\phi$ & $\rho_4\cos\theta$ & $\rho_4\sin\theta$  & $\rho_2$ & $\rho_5$\\
\hline\hline
C1 &$\frac{1}{32} \left(-5-3 \sqrt{17}\right)$ & $-\frac{1}{16} \sqrt{\frac{1}{2} \left(95-7 \sqrt{17}\right)}$ &$
   \frac{1}{64} \left(-23-\sqrt{17}\right)$ & $\frac{1}{32} \sqrt{\frac{1}{2} \left(95-7 \sqrt{17}\right)}$ &$
   \frac{1}{8} \left(\sqrt{17}-1\right)$ & $\frac{1}{32} \left(5+3 \sqrt{17}\right)$ \\
C2& $\frac{1}{32} \left(-5-3 \sqrt{17}\right) $& $\frac{1}{16} \sqrt{\frac{1}{2} \left(95-7 \sqrt{17}\right)}$ &$
   \frac{1}{64} \left(-23-\sqrt{17}\right)$ & $-\frac{1}{32} \sqrt{\frac{1}{2} \left(95-7 \sqrt{17}\right)}
   $& $\frac{1}{8} \left(\sqrt{17}-1\right)$ & $\frac{1}{32} \left(5+3 \sqrt{17}\right)$ \\
C3 &$\frac{1}{32} \left(5-3 \sqrt{17}\right)$ & $-\frac{1}{16} \sqrt{\frac{1}{2} \left(95+7 \sqrt{17}\right)}$ &
   $\frac{1}{64} \left(23-\sqrt{17}\right)$ & $\frac{1}{32} \sqrt{\frac{1}{2} \left(95+7 \sqrt{17}\right)}$ &$
   \frac{1}{8} \left(1+\sqrt{17}\right)$ &$ \frac{1}{32} \left(3 \sqrt{17}-5\right)$ \\
C4& $\frac{1}{32} \left(5-3 \sqrt{17}\right)$ & $\frac{1}{16} \sqrt{\frac{1}{2} \left(95+7 \sqrt{17}\right)}$ &$
   \frac{1}{64} \left(23-\sqrt{17}\right)$ & $-\frac{1}{32} \sqrt{\frac{1}{2} \left(95+7 \sqrt{17}\right)}$ &$
   \frac{1}{8} \left(1+\sqrt{17}\right)$ &$ \frac{1}{32} \left(3 \sqrt{17}-5\right)$ \\
C5& $\frac{1}{32} \left(3 \sqrt{17}-5\right)$ &$ -\frac{1}{16} \sqrt{\frac{1}{2} \left(95+7 \sqrt{17}\right)}$ &$
   \frac{1}{64} \left(\sqrt{17}-23\right)$ &$ \frac{1}{32} \sqrt{\frac{1}{2} \left(95+7 \sqrt{17}\right)} $&$
   \frac{1}{8} \left(-1-\sqrt{17}\right)$ &$ \frac{1}{32} \left(5-3 \sqrt{17}\right)$ \\
C6&$ \frac{1}{32} \left(3 \sqrt{17}-5\right)$ &$ \frac{1}{16} \sqrt{\frac{1}{2} \left(95+7 \sqrt{17}\right)}$ &
 $  \frac{1}{64} \left(\sqrt{17}-23\right)$ &$ -\frac{1}{32} \sqrt{\frac{1}{2} \left(95+7 \sqrt{17}\right)}$ &$
   \frac{1}{8} \left(-1-\sqrt{17}\right)$ &$ \frac{1}{32} \left(5-3 \sqrt{17}\right)$ \\
C7&$ \frac{1}{32} \left(5+3 \sqrt{17}\right) $&$ -\frac{1}{16} \sqrt{\frac{1}{2} \left(95-7 \sqrt{17}\right)}$ &$
   \frac{1}{64} \left(23+\sqrt{17}\right)$ &$ \frac{1}{32} \sqrt{\frac{1}{2} \left(95-7 \sqrt{17}\right)}$ &$
   \frac{1}{8} \left(1-\sqrt{17}\right)$ &$ \frac{1}{32} \left(-5-3 \sqrt{17}\right) $\\
C8& $\frac{1}{32} \left(5+3 \sqrt{17}\right)$ & $\frac{1}{16} \sqrt{\frac{1}{2} \left(95-7 \sqrt{17}\right)}$ &
   $\frac{1}{64} \left(23+\sqrt{17}\right)$ &$ -\frac{1}{32} \sqrt{\frac{1}{2} \left(95-7 \sqrt{17}\right)}$ &$
   \frac{1}{8} \left(1-\sqrt{17}\right)$ &$ \frac{1}{32} \left(-5-3 \sqrt{17}\right)$ \\
 D1&$-\frac{3}{4}$ &$ -\frac{\sqrt{15}}{8}$ &$ -\frac{3}{16}$ & $\frac{\sqrt{15}}{16}$ &$ \frac{1}{8}$ & $\frac{3}{4}$ \\
 D2&$-\frac{3}{4}$ & $\frac{\sqrt{15}}{8}$ & $-\frac{3}{16}$ &$ -\frac{\sqrt{15}}{16}$ & $\frac{1}{8}$ &$ \frac{3}{4}$ \\
D3& $\frac{3}{4}$ & $-\frac{\sqrt{15}}{8}$ & $\frac{3}{16}$ & $\frac{\sqrt{15}}{16}$ &$ -\frac{1}{8}$ &$ -\frac{3}{4}$ \\
D4 & $\frac{3}{4}$ & $\frac{\sqrt{15}}{8}$ & $\frac{3}{16}$ & $-\frac{\sqrt{15}}{16}$ & $-\frac{1}{8}$ & $-\frac{3}{4}$ \\
\hline
\end{tabular}
}
\caption{Solutions of equations \eqref{uni1}-\eqref{uni6} at $N=2$; we omit A1.}
\label{n2sol}
\end{table}

\section{Mapping of nonmixing solutions}
\begin{figure}[t]
\centering
\includegraphics[scale=1.2]{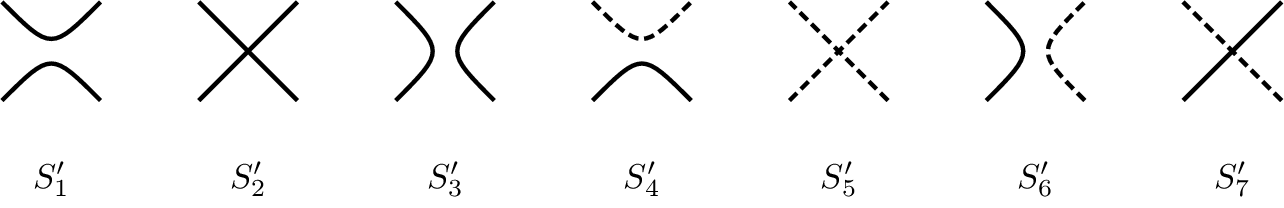}
\caption{Scattering processes for a vector particle multiplet (continuous lines) and a scalar particle (dashed lines).}
\label{vi}
\end{figure}

Equation (\ref{decoup4}) shows that the solutions with $\rho_4=\rho_5=0$ also have $\rho_9=0$, and then $S_4=S_5=S_6=S_9=0$. Figure~\ref{cpampl} shows that the vanishing of these amplitudes eliminates the mixing of indices coming from different particles, and for this reason we refer to this type of solutions as "nonmixing". We now show how, through a change of basis, these nonmixing solutions can all be expressed as those of a system consisting of an $O(N^2-1)$ vector and a scalar that are decoupled. The amplitudes for such a system, in which the scalar and the vector in general interact \cite{DL_vector_scalar}, are shown in figure~\ref{vi} and take the form
\begin{align}
&S'_1=S'^*_3\equiv \rho'_1e^{i\phi'},\\
&S'_2=S'^*_2\equiv \rho'_2,\\
&S'_4=S'^*_6\equiv\rho'_4 e^{i \theta'},\\
&S'_5=S'^*_5\equiv\rho'_5,\\
&S'_7=S'^*_7\equiv \rho'_7\,.
\end{align}
The change of basis that we perform in the $CP^{N-1}$ model is
\begin{equation}
|\Phi_\mu\rangle=\begin{cases}
&|\Phi_0\rangle = \frac{1}{\sqrt{N}}\sum\limits_{a=1}^N |aa\rangle\,,\\
&\frac{1+i}{2}|ab\rangle+\frac{1-i}{2}|ba\rangle\,, \, \, \, \, \mu=ab\,, \, \, \, \,\,\,\,\, a\neq b\,,\\
&\frac{1}{\sqrt{k(k+1)}}\left(\sum\limits_{j=1}^k|jj\rangle-k|(k+1)(k+1)\rangle\right) \,, \, \, \, \, \mu=kk\,, \, \,\,\,\,\, k=1,\dots,N-1,
\end{cases}
\label{map}
\end{equation}
with $\langle\Phi_\mu|\Phi_\nu\rangle=\delta_{\mu\nu}$, and the trace mode $\Phi_0$ being the scalar of the vector-scalar system. The scattering matrix for the non-mixing case of the $CP^{N-1}$ model can now be expressed as
\begin{equation}
\begin{split}
S_{\mu, \nu}^{\rho, \sigma} &= \left (S_1^\prime \delta_{\mu, \nu} \delta^{\rho, \sigma} + S_2^\prime \delta_{\mu}^{\rho}\delta_{\nu}^{\sigma} + S_3^\prime \delta_{\mu}^{\sigma} \delta_{\nu}^{\rho} \right ) \bar{\delta}_{\mu}^0\bar{\delta}_{\nu}^0\bar{\delta}_{0}^{\rho}\bar{\delta}_{0}^{\sigma} + S_4^\prime (\delta_{\mu, \nu} \delta_{0}^\rho \delta_{0}^\sigma\bar{\delta}_\mu^0\bar{\delta}_\nu^0 + \delta_{\mu}^0 \delta_\nu^0 \delta^{\rho, \sigma}\bar{\delta}_0^\rho\bar{\delta}_0^\sigma ) \\ 
&\quad + S_5^\prime \delta_{\mu}^0 \delta_\nu^0 \delta_0^\rho \delta_0^\sigma + S_6^\prime (\delta_{\mu}^\sigma \delta_\nu^0 \delta_0^\rho\bar{\delta}_\mu^0\bar{\delta}_0^\sigma + \delta_\mu^0 \delta_0^\sigma \delta_\nu^\rho\bar{\delta}_\nu^0\bar{\delta}_0^\rho) + S_7^\prime ( \delta_\mu^\rho \delta_\nu^0 \delta_0^\sigma\bar{\delta}_\mu^0\bar{\delta}_0^\rho + \delta_\mu^0 \delta_0^\rho \delta_\nu^\sigma\bar{\delta}_\nu^0\bar{\delta}_0^\sigma)\,,
\end{split}
\end{equation}
where $\bar{\delta}_{\mu}^{\nu}\equiv 1- \delta_{\mu}^{\nu}$, and
\begin{align}
S_1^\prime &= \langle \Phi_\nu \Phi_\nu|\mathbb{S}| \Phi_\mu \Phi_\mu\rangle = S_1, \label{smap1}\\
S_2^\prime &= \langle \Phi_\mu \Phi_\nu|\mathbb{S}| \Phi_\mu \Phi_\nu\rangle = S_2, \\
S_3^\prime &= \langle \Phi_\nu \Phi_\mu|\mathbb{S}| \Phi_\mu \Phi_\nu\rangle = S_3, \\
S_4^\prime &= \langle \Phi_0 \Phi_0|\mathbb{S}| \Phi_\mu \Phi_\mu\rangle = \langle \Phi_\nu \Phi_\nu|\mathbb{S}| \Phi_0 \Phi_0 \rangle = S_1+NS_{11}\\
S_5^\prime &= \langle \Phi_0 \Phi_0|\mathbb{S}| \Phi_0 \Phi_0 \rangle = S_1+S_2+S_3+2N(S_7+S_8)+N^2S_{10}+2NS_{11}, \\
S_6^\prime &= \langle  \Phi_\mu \Phi_0|\mathbb{S}| \Phi_0 \Phi_\mu\rangle = \langle \Phi_0 \Phi_\nu|\mathbb{S}| \Phi_\nu \Phi_0\rangle  = S_3+NS_8, \\
S_7^\prime &= \langle \Phi_0 \Phi_\mu|\mathbb{S}| \Phi_0 \Phi_\mu\rangle = \langle \Phi_\nu \Phi_0|\mathbb{S}| \Phi_\nu \Phi_0\rangle  = S_2+NS_7\,. \label{smap7}
\end{align}

\noindent Using the trace decoupling equations \eqref{decoup1}-\eqref{decoup5} the relations \eqref{smap1}-\eqref{smap7} reduce to
\begin{equation}
S'_1=S_1 \, \, , \, \, S'_2=S_2 \, \, , \, \, S'_3=S_3 \, \, , \, \, S'_4=S'_6=0 \, \, , \, \, S'_5=S'_7=S_0\,,
\end{equation} 
which exhibit the decoupling between the vector and the scalar (recall that $S_0=\pm1$). Table~\ref{cpvs} gives the explicit form of the $CP^{N-1}$ nonmixing solutions in terms of the vector-scalar amplitudes. One sees, in particular, that solutions A1a$_{\pm}$ and A1b$_{\mp}$ only differ for the nature of the decoupled scalar (fermionic or bosonic).

\begin{table}[h!]
\centering
\begin{tabular}{|c|c|c|c|c|c|c|c|}
\hline
Solution & $N^2-1$ &$\rho'_1$& $\rho'_2$ & $\cos\phi'$ & $\rho'_4$ &$\rho'_5$& $\rho'_7$\\
\hline\hline
A1a${_\pm}$ &$\mathbb{R}$ & $0$ & $S_0$ & $-$ & $0$ & $S_0$ & $S_0$\\
A1b${_\pm}$ &$\mathbb{R}$ & $0$ & $-S_0$ & $-$ & $0$ & $S_0$ & $S_0$\\
A2$_\pm$  &$\left[-2,2\right]$ & $1$ & $0$ & $(\pm)\frac{1}{2}\sqrt{3-N^2}$ & $0$ &$S_0$ & $S_0$\\
A3$_\pm$ & $2$ &$\sqrt{1-\rho_2^2}$ & $[-1,1]$ & $0$ & $0$ & $S_0$ & $S_0$\\
\hline
\end{tabular}
\caption{Nonmixing solutions of the $CP^{N-1}$ model in terms of the amplitudes of the vector-scalar system. Signs in parenthesis are both allowed, and $S_0=\pm1$.}
\label{cpvs}
\end{table}


\end{document}